\documentclass[aps,prapplied,longbibliography,10pt,twocolumn]{revtex4-2}

\usepackage[english]{babel}
\usepackage{subcaption}
\usepackage{amsmath}
\usepackage{graphicx}
\usepackage{multirow}
\usepackage{subcaption}
\usepackage[colorlinks=true, allcolors=blue]{hyperref}
\usepackage{siunitx}

\DeclareSIUnit\sqrthz{\ensuremath{\sqrt{\text{Hz}}}}
\DeclareSIUnit[per-mode = symbol]\Hzasd{\Hz\per\sqrthz}
\DeclareSIUnit[per-mode = symbol]\masd{\m\per\sqrthz}
\DeclareSIUnit[per-mode = symbol]\fmasd{\femto\m\per\sqrthz}
\DeclareSIUnit\month{month}

\begin{document}

\title{Sensitivity and control of a 6-axis fused-silica seismometer}
\date{\today}

\author{Jiri Smetana}\email{gsmetana@star.sr.bham.ac.uk}
\affiliation{Institute for Gravitational Wave Astronomy, School of Physics and Astronomy, University of Birmingham, Birmingham B15 2TT, United Kingdom}


\author{Amit Singh Ubhi}
\affiliation{Institute for Gravitational Wave Astronomy, School of Physics and Astronomy, University of Birmingham, Birmingham B15 2TT, United Kingdom}

\author{Emilia Chick}
\affiliation{Institute for Gravitational Wave Astronomy, School of Physics and Astronomy, University of Birmingham, Birmingham B15 2TT, United Kingdom}

\author{Leonid Prokhorov}
\affiliation{Institute for Gravitational Wave Astronomy, School of Physics and Astronomy, University of Birmingham, Birmingham B15 2TT, United Kingdom}

\author{John Bryant}
\affiliation{Institute for Gravitational Wave Astronomy, School of Physics and Astronomy, University of Birmingham, Birmingham B15 2TT, United Kingdom}

\author{Artemiy Dmitriev}
\affiliation{Institute for Gravitational Wave Astronomy, School of Physics and Astronomy, University of Birmingham, Birmingham B15 2TT, United Kingdom}

\author{Alex Gill}
\affiliation{Institute for Gravitational Wave Astronomy, School of Physics and Astronomy, University of Birmingham, Birmingham B15 2TT, United Kingdom}

\author{Lari Koponen}
\affiliation{Institute for Gravitational Wave Astronomy, School of Physics and Astronomy, University of Birmingham, Birmingham B15 2TT, United Kingdom}

\author{Haixing Miao}
\affiliation{Department of Physics, Tsinghua University, Beijing, China 100084}

\author{Alan V. Cumming}
\affiliation{Institute for Gravitational Wave Research, School of Physics and Astronomy, University of Glasgow, Glasgow G12 8QQ, United Kingdom}

\author{Giles Hammond}
\affiliation{Institute for Gravitational Wave Research, School of Physics and Astronomy, University of Glasgow, Glasgow G12 8QQ, United Kingdom}

\author{Valery Frolov}
\affiliation{LIGO Livingston Observatory, Livingston, LA 70754, USA}

\author{Richard Mittleman}
\affiliation{LIGO Laboratory, Massachusetts Institute of Technology, Cambridge, Massachusetts 02139, USA}

\author{Peter Fritchel}
\affiliation{LIGO Laboratory, Massachusetts Institute of Technology, Cambridge, Massachusetts 02139, USA}

\author{Denis Martynov}
\affiliation{Institute for Gravitational Wave Astronomy, School of Physics and Astronomy, University of Birmingham, Birmingham B15 2TT, United Kingdom}

\begin{abstract}
We present a pair of seismometers capable of measurement in all six axes of rigid motion. The vacuum-compatible devices implement compact interferometric displacement sensors to surpass the sensitivity of typical electrical readout schemes. Together with the capability to subtract the sensitivity-limiting coupling of ground tilt into horizontal motion, our seismometers can widen the sensing band towards mHz frequencies. This has notable applications across a range of fields requiring access to low-frequency signals, such as seismology and climate research. We particularly highlight their potential application in gravitational-wave observatories (LIGO) in improving their observation capability of intermediate-mass black holes ($\sim 1000\,M_\odot$). The sensors are based on a near-monolithic fused-silica design consisting of a fused-silica mass and fibre, showing improved stability and robustness to tilt drifts, alignment, and control compared to all-metal or mixed metal-silica designs. We demonstrate tilt sensitivity that surpasses the best commercial alternatives in a significantly reduced footprint compared to our previous iterations of these sensors.
\end{abstract}

\maketitle

\section{Introduction}
\label{sec:intro}

Inertial sensors are devices that measure motion relative to an inertial frame of reference and have a vast range of applications~\cite{InertialTrends}. Such measurements are useful for tracking the motion, orientation, and position of objects, which makes them an essential component of inertial navigation~\cite{InertialNav1,InertialNav2}. Significant advances have been made in recent years in miniaturising the associated technology and deploying it on an industrial scale. Accelerometers based on micro-electro-mechanical systems (MEMS)~\cite{MEMS} see widespread use in modern smartphones, drones and other compact electronic devices. Recent developments in optical sensing are reflected in various optical gyroscopes~\cite{OpticalGyro1,OpticalGyro2}, and devices implementing whispering gallery mode resonators~\cite{WhisperingGallery}. Highly pure materials such as fused silica/quartz have been used to make compact single-axis sensors~\cite{SilicaAccel1,SilicaAccel2}. Larger and more robust inertial sensors are typically used within the field of geophysics as seismometers or geophones to measure the ambient~\cite{AmbientSeismic} and transient~\cite{TransientSeismic} vibrations of the ground to study earthquakes and probe the structure of the Earth's crust.

However, even more precise sensors are required for fundamental physics research, including observation of gravitational waves. In this case inertial sensors are used within a broader feedback scheme to sense and control the motion of platforms and larger devices---the technique of inertial isolation. Inertial isolation can also be seen as an essential enabling technology for quantum sensing and computing development, and applications in industrial manufacturing processes.

As a general rule, inertial sensors typically take the form of a mass-spring system (or equivalent). The mass acts as the inertial reference above the system's mechanical resonance whose relative motion yields a measure of the environmental vibrations. The test-mass motion below the mechanical resonance becomes common with the environment degrading the sensor response with a characteristic $f^2$ dependence. Typical sensors are mechanically constrained to allow motion along a single axis of motion (perhaps up to three), which makes them mechanically simple and robust but they consequently suffer from cross-couplings between different degrees of freedom (DoF; see Section~\ref{sec:ligo}). We focus on three crucial deficiencies of such sensors. These are (i) the choice of readout scheme, (ii) the inability to decouple different DoFs, and (iii) their degrading low-frequency response.

In this paper, we present the design and performance of two six-axis inertial sensors based on ideas laid out in Ref.~\cite{6DIdea}. The purpose of these devices is to address the above limitations through (i) taking advantage of compact interferometric sensors with sub-picometer displacement sensitivity, (ii) creating an unconstrained system with full six-axis freedom of motion, and (iii) achieving low resonant frequencies along key DoFs. The ultimate goal is to provide better sensitivity across the full sensing band and to extend this band towards lower frequencies. The final designs and performance presented here follow extensive prior development of numerous metal precursors~\cite{C6DProto,6D1,6D2}. We enhanced the sensitivity and robustness (in terms of both mechanical and measurement stability) of the seismometers by (i) utilising fused silica instead of aluminium to suppress drifts, (ii) making the devices compact to fit into LIGO-type inertial isolation platforms, (iii) developing compact interferometers with deep-frequency modulation readout, and (iv) introducing `magnetoguns' (see Section~\ref{sec:sensing_actuation}) to correct the angular position of the mass.

These devices are versatile and thus can benefit a multitude of disparate fields of research. The sensor can be used directly as a seismometer, where it can detect low-frequency signals and thus obtain more complete information about seismic transients, as well as to provide insight into low-frequency seismic phenomena. The importance of low frequency seismic information has been demonstrated in earthquake analysis~\cite{SeismicEarthquake}, hydrocarbon surveys~\cite{SeismicPassive}, and climate research~\cite{SeismicClimate}. In all of these applications, the sensor would be effective due to its ability to decouple tilt from horizontal motion and its high sensitivity through the use of interferometric sensors. Better sensitivity can also reduce the amplitude of acoustic excitations generated during active seismic surveys, which would be less disruptive to local wildlife~\cite{SeismicWildlife}.

Our inertial sensors can be integrated into actively controlled platforms and provide a vibration-free environment for general use. Often only a few DoFs are directly important for the given application. However, cross-couplings between the different DoFs inevitably lead to the most effective solution being to control all six DoFs regardless. Our sensor would be beneficial in such applications due to its high sensitivity and ability to measure all six DoFs in a single device. Most prominently for us is its use in gravitational wave (GW) detectors~\cite{LIGOSeismicStrategy,VirgoSuperattenuator,KAGRASeismic,ETEST} (see Section~\ref{sec:ligo}). Additionally, our sensor can be applied to suppress vibrations of cryogenic systems~\cite{TOBA,CryoCavity}, electron microscopes~\cite{Microscopes} and used as a test-bed for space-borne instruments~\cite{SpaceTesting}. Outside of research applications, the sensors can be used in ultra-precision manufacturing~\cite{Manufacturing} and photolithography~\cite{Photolith1,Photolith2}.

Finally, the sensors we developed can be used in the role of a torsion balance. As part of our strict requirements for low-frequency sensitivity, we achieve low eigenfrequencies in the angular DoFs. This is achieved by adopting a design based on a torsion pendulum. Ultimately, we reach a level of sensitivity in the torsional DoF that is competitive with contemporary torsion balance experiments. Torsion balances have a strong history in gravity measurements with recent examples including Bosonic dark matter searches~\cite{DarkMatter}, and Newtonian noise measurement~\cite{Newtonian}. We have recently adapted the precursor 6D sensor~\cite{6D2} to investigate semi-classical gravity models~\cite{SemiClassical1,SemiClassical2,SemiClassical3}, with results currently in preparation. It is also possible to apply such a device in air-borne gravity gradiometry~\cite{Gravity1,Gravity2} and short-range force measurements investigating the Casimir effect in meta-materials~\cite{Casimir1,Casimir2,Casimir3}.

\section{LIGO Application}
\label{sec:ligo}

\begin{figure*}[t]
    \centering
    \includegraphics[width=0.95\textwidth]{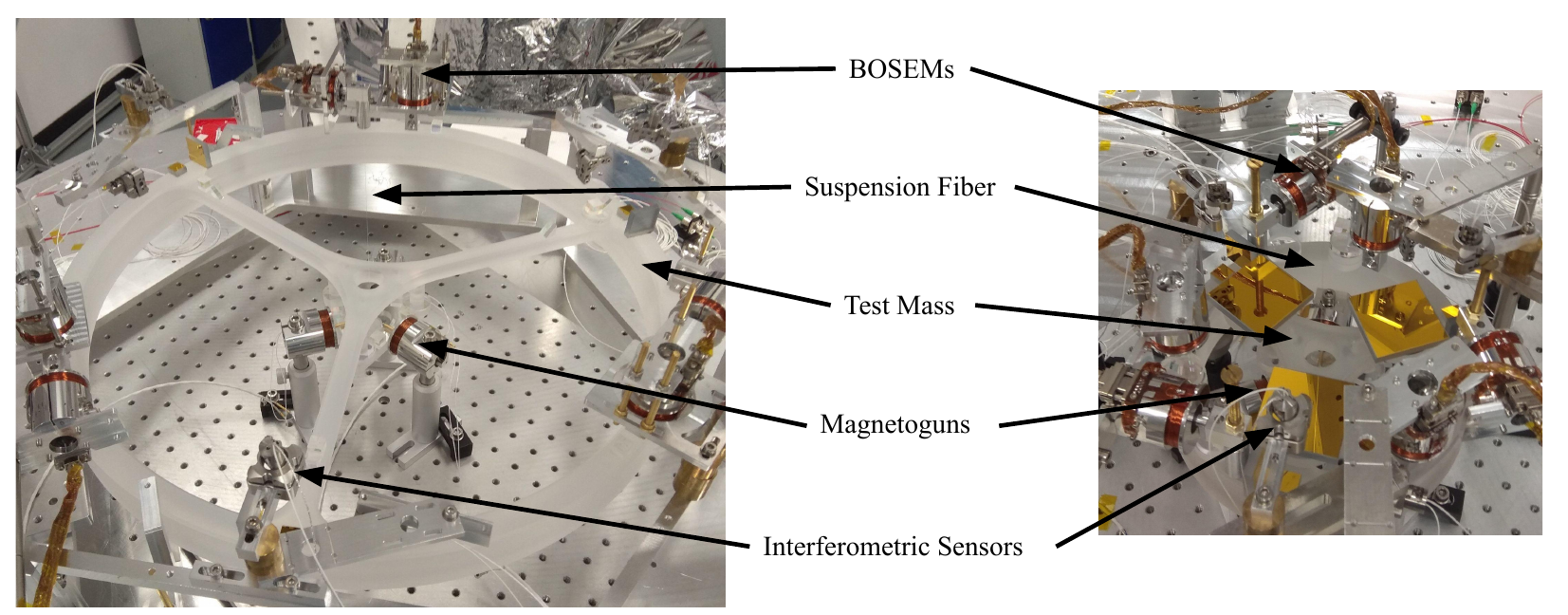}
    \caption{Annotated picture of the C-6D and M-6D seismometers with key components of the setup: interferometers, coil-magnet actuators (BOSEMs), and magnetoguns. Magnetoguns on M-6D are mostly obscured from view by the gold coated ballast masses. For a sense of scale, the outer diameter of C-6D and M-6D rings are \SI{50}{\cm} and \SI{20}{\cm}, respectively.}
    \label{fig:scheme}
\end{figure*}

The long-standing focus of our sensor development has been on the challenges faced by the gravitational-wave (GW) community, providing a compelling case study of the issues our sensors are designed to address. Here, we recapitulate the principal points raised in Ref.~\cite{6D2} for better context.

The Advanced LIGO~\cite{AdvLIGO} and Advanced Virgo~\cite{AdvVirgo} detectors have successfully yield dozens of GW sources~\cite{Catalogue1,Catalogue2,Catalogue3} since the first GW transient detection in 2015~\cite{BBHDetection}. However, the duty cycle of the detectors, as well as their sensitivity below 25 Hz~\cite{SensGW,LIGOSensO3}, is significantly limited by ground vibrations (specifically, controls noise from the feedback loops used to control the position of the primary optics). The reduced duty cycle leads to the loss of otherwise detectable sources and the narrower detection band~\cite{5Hz} limits the accumulation of signal-to-noise ratio. This degrades the parameter estimation and sky localisation, and limits the early warning time~\cite{EarlyWarning} for follow-up and multi-messenger observations~\cite{MultiMessenger}. Widening the band also increases the prospects to detect and study intermediate-mass black holes~\cite{IMBHDetection}.

The sharp suppression of seismic noise in Advanced LIGO is partly achieved by the quadruple pendulum~\cite{LIGOQuad} suspension. This passive approach is augmented by a complex network of seismic sensors and electromagnetic actuators acting in feedback and feedforward to actively suppress motion of the Internal Seismic Isolation (ISI) platform~\cite{LIGOSeismicStrategy,ISI1} which hosts the quadruple suspension. This platform is controlled in all six DoFs over the 100 mHz--40 Hz band~\cite{ISI2} with inertial sensing performed by a range of commercial geophones and seismometers such as the Geotech GS-13 and the Trillium 240. The insufficient sensitivity of these inertial sensors, particularly below their typical \SI{1}{\Hz} resonance, is a key limitation and a crucial problem that must be solved in order for the next generation detectors, Cosmic Explorer~\cite{CE1,CE2} and Einstein Telescope~\cite{ET1,ET2}, to reach their design sensitivity (see e.g. Ref.~\cite{C6DProto}).

Commercial sensors are limited by the noise of their inductive/capacitive readout exacerbated by the degrading response below \SI{1}{\Hz}. Horizontal single-axis sensors further suffer from tilt-horizontal cross-couplings. Ground tilt around the orthogonal horizontal axis couples into the readout with a factor of $g / \omega^2$~\cite{Tilt2Hor}, where $\omega$ is the angular frequency, which masks the true horizontal signal at low frequencies. One solution is to make an independent estimate of the ground tilt with two spatially separated vertical sensors and subtract it from the horizontal sensor. This method is limited by the relatively short, metre-scale coherence length of the ground tilt and the space constraints on the typical isolated platform. It also relies on multiple stages of coherent subtraction, up to a common-mode rejection ratio of 1000, which is highly susceptible to calibration errors and systematics.

A wide range of custom inertial sensors have been demonstrated or proposed in the literature. Sensors with minimal tilt coupling have been proposed~\cite{TiltFree} to bypass the cross-coupling effect by, for example, investigating novel geometries and using multi-stage mechanical systems that offer negligible coupling to ground tilt. In the other direction, dedicated rotation sensors~\cite{BRS1,BRS2,CRS} have been developed, consisting of beams or cylinders with large moment of inertia, which offer high sensitive in tilt but negligible translational coupling. These sensors can be used to subtract the tilt signals that pollute conventional horizontal sensors. Improved sensors based around optical readout have also been proposed~\cite{OpticalAccel1,OpticalAccel2,OpticalAccel3,HoQIGeophone,InertSens1,InertSens2}. These sensors are either variations upon the conventional single-axis design, or make sure of some novel geometry but, importantly, they demonstrate the strong application of optical (often interferometric) sensors in reaching substantial improvements in read-out sensitivity. Our focus has been on the development of a single six-axis seismometer that overcomes all of the key limitations simultaneously in a single device.

\section{Mechanical Design}
\label{sec:mech_design}

The two sensors we present here, Compact 6D (C-6D) and Mini 6D (M-6D), fundamentally share the same design principles and the same technologies but differ in size. We effectively developed two versions of the same seismometer to investigate the scaling of its sensitivity with size. The core of the seismometer consists of a fused-silica test mass suspended by a single thin fused-silica fibre. The test mass position is measured by an array of interferometric displacement sensors and optical shadow sensors. The sensors measure the mass position relative to the enclosing frame, which holds the sensors and supports the fibre. The test mass position is controlled in feedback by magnet-coil actuators incorporated into the shadow sensor design. Annotated photos of the two sensors are shown in Fig.~\ref{fig:scheme}. The sensing and control scheme is very similar to that shown in Figs~1\&2 in Ref.~\cite{6D2} and the detailed schematic of the interferometric sensor readout can be found in Fig.~1 of Ref.~\cite{Smaract}. The parameters of the mechanical design of both sensors are provided in Table~\ref{tab:properties}. For reference, we define the Z DoF as vertical and co-aligned with the fibre's central axis, the X and Y DoFs are along the two orthogonal horizontal axes passing through the centre of mass (CoM) of the test mass, and RX, RY, RZ are the rotations around the corresponding axes.

\begin{table}[ht]
    \begin{tabular}{|c|c|c|c|}
        \hline
        Property & C-6D & M-6D & Unit\\
        \hline
        TM - Mass & 3.3 & 1.9 & \unit{\kg} \\
        TM - Diameter & 500 & 200 & \unit{\mm} \\
        TM - Thickness & 40 & 40 & \unit{\mm} \\
        TM - MoI (RX,RY) & 0.079 & 0.005 & \unit{\kg \m \squared} \\
        TM - MoI (RZ) & 0.16 & 0.01 & \unit{\kg \m \squared} \\
        Fibre Diameter & 240 & 240 & \unit{\um} \\
        Fibre Length & 520 & 520 & \unit{\mm} \\
        Mean Stress & 715 & 412 & \unit{\MPa} \\
        \hline
    \end{tabular}
    \caption{Parameters of the test mass (TM) and fibre. Mass and moments of inertia (MoI) include the contribution of sensing components and additional ballast masses.}
    \label{tab:properties}
\end{table}

An essential component of our design is the single fused-silica suspension fibre---the same fibre in both tests. This fibre was pulled at the University of Glasgow similarly to the fibres on our precursor designs, using the same technique developed for the Advanced LIGO quadruple suspensions~\cite{FusedSilicaFibres}. We observe the beneficial linear, elastic behaviour of the fibre, which exhibits no detectable hysteresis as witnessed in the RZ torsional motion. We witness this via the fibre returning to its initial position after responding to transient changes in conditions, such as pressure changes and thermal cycling during pump-down. In contrast, this property is notably different in other DoFs, such as tilt, where the mass shifts to a new mean position in response to such transients. This is due to the nonlinear behaviour of other materials, such as epoxy, which strongly influence those DoFs (see below). This, among other properties of fused silica, is a key motivation for our move towards a monolithic fused-silica design, replacing the previous aluminium test mass.

The fused-silica test mass is wheel-shaped with most of the mass concentrated at the edges of the ring to maximise the moment of inertia for the given mass. This allows us to reduce the actuation strength needed to control the linear DoFs and use a thinner fibre. The higher moment of inertia reduces the tilt resonances and thus leads to better tilt sensitivity. Here, we see the benefits of using fused-silica for the mass. In particular, the lower coefficient of thermal expansion ($\sim$\SI{5e-7}{\per\kelvin}) leads to a small sensitivity to thermal gradients, which we see as a decrease in the (RX, RY) tilt drift. For a representative tilt resonance of \SI{8}{\mHz}, the tilt drift is \SI[per-mode=symbol]{2.5}{\milli \radian \per \month} for the fused-silica C-6D as opposed to \SI[per-mode=symbol]{7.3}{\milli \radian \per \month} for the aluminium precursor. The tilt drift of C-6D is limited by the epoxy bonding mechanism (see below) otherwise drift reduction would likely be even higher. Further improvements to the mass stability, particularly to temperature fluctuations and thermal gradients, can be achieved by applying a gold coating to the silica mass to reduce its emissivity. This should not increase the thermal noise as  the dominant thermal noise contribution originating from regions that store large elastic potential energy---the lossy epoxy layer connecting fibre to mass and the effective bending point of the fused silica fibre. Neither region will be gold-coated and hence will not increase the thermal noise.

The fibre is welded at both ends to a fused-silica anchor cone to present a larger surface area and longer lever arm for bonding. The silica welding was also performed by the University of Glasgow. The monolithic fibre-anchor component is connected to the test mass via an intermediate fused-silica interface plate. The interface plate is bonded to the test mass proper via three vertical stacks of fused-silica spacers. These spacers allow for the coarse tuning of the vertical CoM position relative to the effective bending point of the fibre. The tilt resonances are dictated by the elastic restoring force of the bent fibre, and the gravitational restoring force of the compound pendulum. The tilt frequency is given by
\begin{equation}
    f^2 = \frac{mgh + k_{el}}{4\pi^2 I},
    \label{eq:tilt_freq}
\end{equation}
where $h$ is the distance between the centre of mass and the effective lower bending point of the fibre, $I$ is the moment of inertia, and $k_{el}$ is the elastic spring constant of the fibre~\cite{SusThermViolin1,SusThermViolin2}. The latter is given by $k_{el} = \sqrt{m g E I_a}/2$, where $E$ is the Young's modulus of fused-silica and $I_a$ is the second moment of area, given by $I_a = \pi r^4 / 4$ in the case of a circular cross-section fibre.

We can place the centre of mass above the bending point (corresponding to negative $h$ in Eq.~\ref{eq:tilt_freq}) and generate a gravitational anti-spring effect. This reduces the overall angular stiffness analogously to the inverted pendulum, which has a history of use within the GW community~\cite{VirgoSuperattenuator,InvertedPend}. In consequence, we can theoretically reach any resonant frequency regardless of the test mass properties by properly tuning the vertical CoM position. Micron-level fine-tuning can be achieved by adding/removing gram-scale balancing masses. In principle, this means that there is no penalty to tilt sensitivity of M-6D from the mechanical response, despite its smaller size compared to C-6D. The only reduction in sensitivity comes from the larger readout noise due to the latter's factor of 2.5 longer lever arm. In practise, it is challenging to fine-tune the CoM offset for compact test masses. Due to M-6D's significantly smaller moment of inertia, its tilt resonance is approximately 4 times more sensitive to the change in CoM height than C-6D (i.e. $df / dh$ derived from Eq.~\ref{eq:tilt_freq} is 4 times larger), with only a \SI{1}{\um} difference separating a tilt frequency of \SI{10}{\mHz} from instability.

In line with our past observations, the bonding mechanism is crucial in determining the mechanical Q-factors, and for reducing long-term drifts, which arise from stress relaxations and dislocation avalanches~\cite{DislocationAvalanche} concentrated within the bonding material. The latter also gives rise to crackling noise~\cite{Crackling} and non-stationary effects, which we have previously observed in the tilt spectrum below \SI{50}{\mHz}. From past experience, we found that the best bond is achieved with a thin layer of Araldite 2014-2 epoxy (see Ref.~\cite{6D2} for a discussion of the alternatives). In subsequent improvements, we will weld the silica components together into a completely monolithic assembly to eliminate the epoxy interface. Such an approach has been successfully demonstrated on the ultimate suspension stage of the LIGO quadruple pendulum~\cite{FusedSilicaFibres}.

\begin{table}[ht]
    \begin{tabular}{|c|c|c|c|c|}
        \hline
        \multirow{2}{*}{Mode} & \multicolumn{2}{c|}{C-6D} & \multicolumn{2}{c|}{M-6D} \\
        \cline{2-5}
         & Frequency & Q-Factor & Frequency & Q-Factor \\
        \hline
        X & \SI{0.69}{\Hz} & $>10^4$ & \SI{0.69}{\Hz} & $>10^4$ \\
        Y & \SI{0.69}{\Hz} & $>10^4$ & \SI{0.69}{\Hz} & $>10^4$ \\
        Z & \SI{6.9}{\Hz} & $>10^4$ & \SI{9.5}{\Hz} & $>10^4$ \\
        RX & \SI{29}{\mHz} & 120 & \SI{68}{\mHz} & 170 \\
        RY & \SI{5.9}{\mHz} & 5 & \SI{6.9}{\mHz} & 2 \\
        RZ & \SI{1.48}{\mHz} & $>10^4$ & \SI{5.9}{\mHz} & $>10^4$ \\
        \hline
    \end{tabular}
    \caption{Resonances and associated Q-factors for all six DoFs for both C-6D and M-6D sensors.}
    \label{tab:modes}
\end{table}

The resonant frequencies and Q-factors of all DoFs of the final sensor configurations are given in Table~\ref{tab:modes}. The X and Y resonances for both sensors are simply given by the pendulum mode given by the fibre length in Table~\ref{tab:properties}. The fibre bending stiffness does not have a significant effect on this mode due to the large gravitational dilution factor~\cite{SusThermViolin1,SusThermViolin2}. The Z resonance is given as a simple mass-spring system dictated by the effective stiffness of the fibre in tension. The Z DoF is relatively stiff in comparison with widely available singe-axis vertical sensors. This is due to us not implementing dedicated vertical isolation in the current iteration of the sensor. Our purpose is to demonstrate the performance of the sensor in the key horizontal and tilt DoFs, which makes vertical performance less important. In fact, a stiff Z DoF allows us to obtain an independent estimator for the nonlinear readout noise floor (see below) and thus serves an important purpose. In a future upgrade, we will reduce the Z resonance through the use of blade springs. The RZ resonance is given by the torsional stiffness of the fibre and the RZ moment of inertia. The RX and RY resonances are chosen by us as compromise between sensitivity and stability via the vertical CoM fine-tuning. As shown in Table~\ref{tab:modes}, the RX and RY resonances are not the same, which is due to unplanned ellipticity of the suspension fibre, which leads to $k_{el,x} \neq k_{el,y}$ in Eq.~\ref{eq:tilt_freq}. The observed splitting suggests approximately 10\% discrepancy between the effective diameter in the two orthogonal axes. The CoM tuning cannot affect the two resonances independently, which means we cannot eliminate the resonance splitting via independent tuning. In practice, this adds a lower limit on the larger of the two tilt resonances if both DoFs are to remain stable. This frequency splitting is reduced by the moment of inertia, making C-6D more resistant to this splitting by an additional factor of 2.4 compared to M-6D. This further highlights some of the compromises made by reducing the size of the test mass, and provides a compelling reason for obtaining the highest quality suspension fibres. It is possible to compensate for the increased difficulties of a smaller test mass by raising the overall mass or by reducing the thickness of the suspension fibre.

Many Q-factors in Table~\ref{tab:modes} are only given as a lower bound due to the extremely low mechanical loss of silica fibres (Q-factors in excess of \num{1e7} have been demonstrated with fused-silica pendulums~\cite{LossSilica}). The RX/RY Q-factors are significantly lower due to the dominant contribution from the epoxy layer that primarily experiences deformations in reaction to test mass tilt. All DoFs additionally experience eddy current damping arising from magnets attached to the mass as part of the actuator scheme. This is a variable effect depending on the control force exerted by DC stabilisation servos, which is discussed in more detail in the next section. We calculate that the eddy current damping in the absence of stabilisation is unlikely to limit the sensitivity in any degrees of freedom. Significant improvements in other noise sources, such as readout and actuation noise will have to be addressed before the eddy-current-limited thermal noise becomes an issue. However, there is potential to mitigate this in the future by replacing the aluminium body of the actuators with peek counterparts if necessary.

\section{Sensing and Actuation}
\label{sec:sensing_actuation}

\begin{figure*}
    \centering
    \begin{subfigure}{0.49\linewidth}
        \centering
        \includegraphics[width=\linewidth]{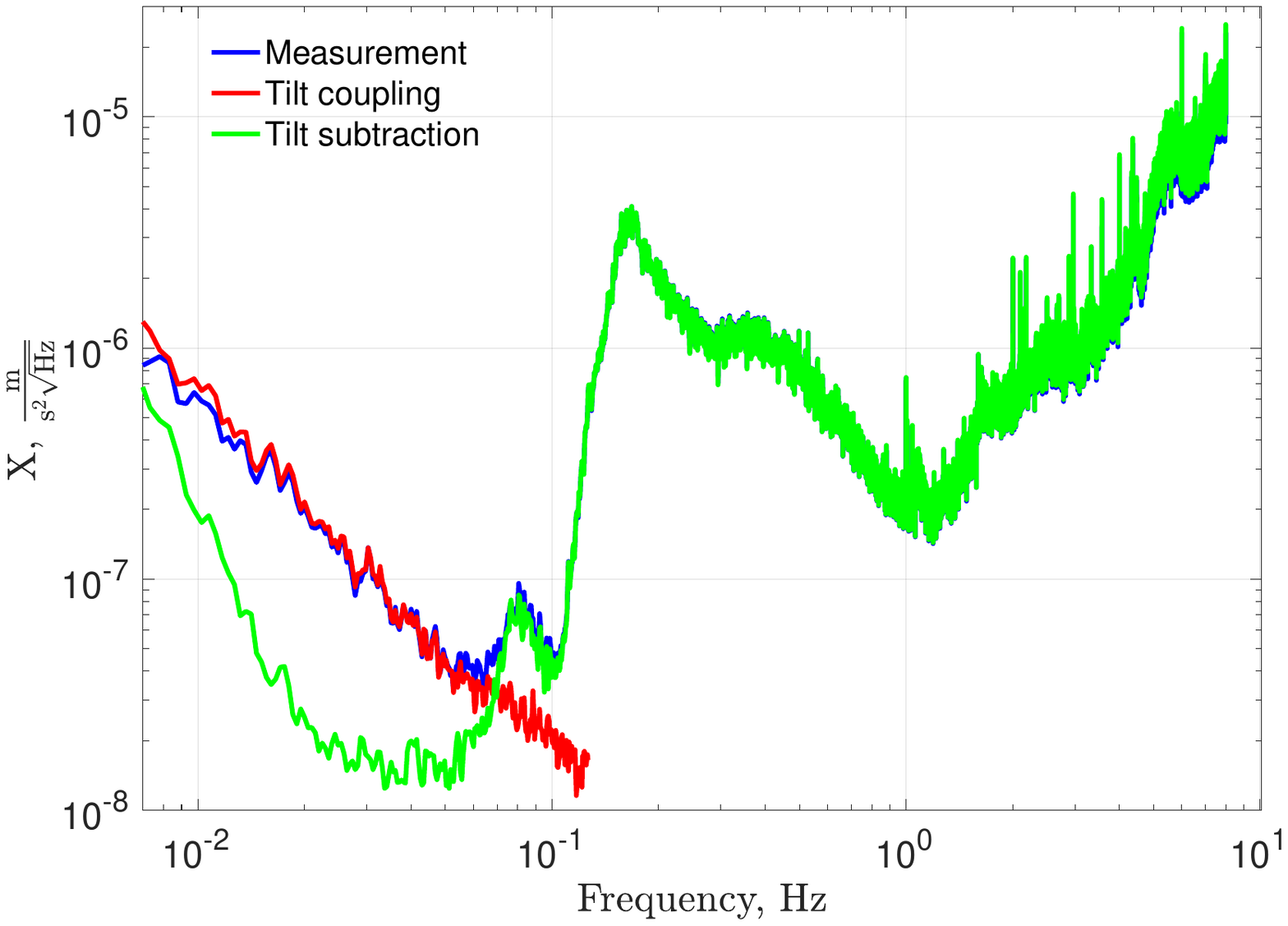}
        \caption{Horizontal acceleration}
        \label{subfig:sens_x}
    \end{subfigure}
    \begin{subfigure}{0.49\linewidth}
        \centering
        \includegraphics[width=\linewidth]{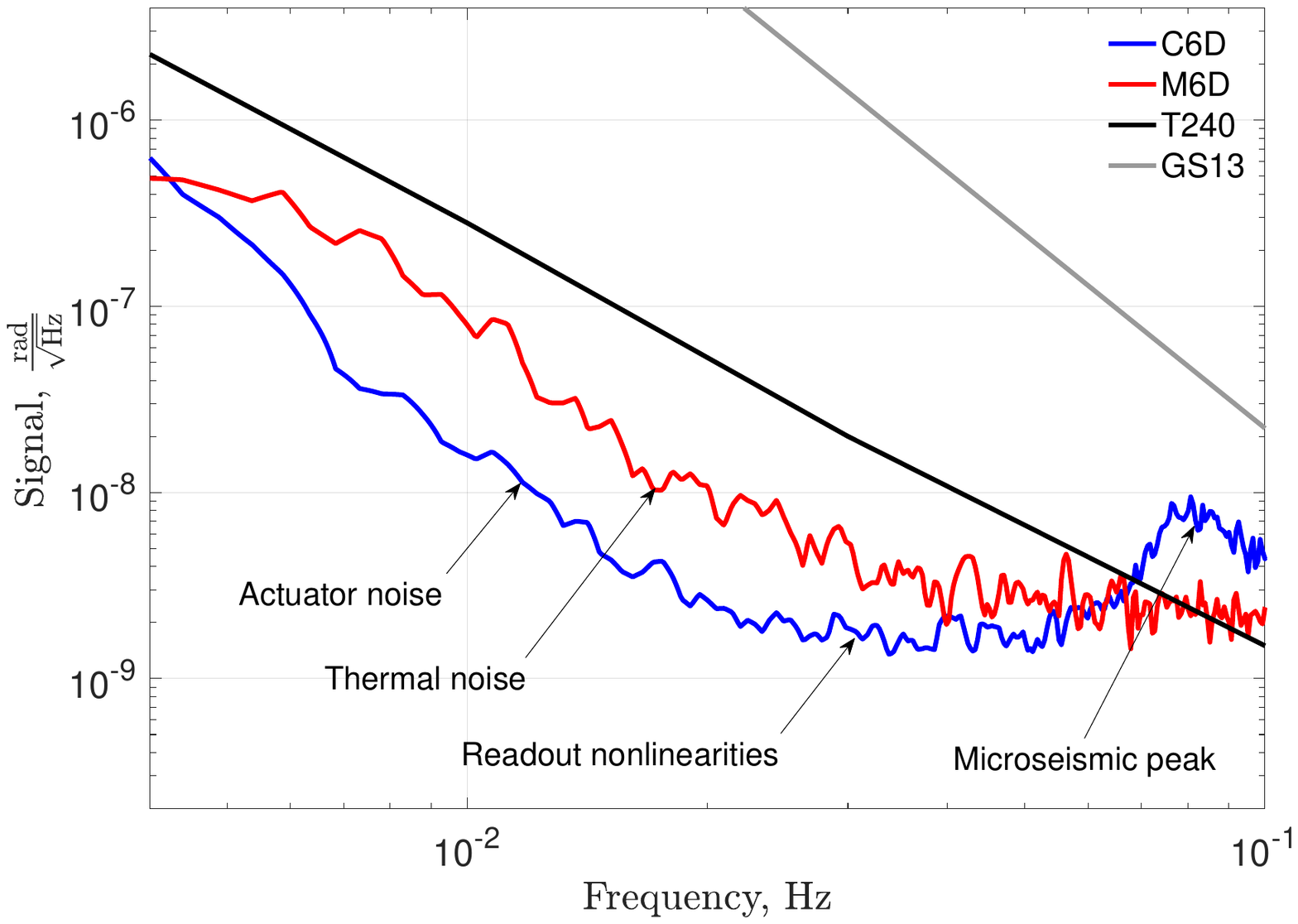}
        \caption{Tilt sensitivity}
        \label{subfig:sens_tilt}
    \end{subfigure}

    \caption{Sensitivity of the two seismometers in key degrees of freedom. Panel (a) shows horizontal acceleration representative of both C-6D and M-6D performance in both X and Y and demonstrates the effective removal of tilt coupled motion by up to an order of magnitude below \SI{100}{\mHz}. The tilt subtraction demonstrated here is for the C-6D sensor. Panel (b) shows C-6D and M-6D tilt sensitivity compared to key commercial alternatives with dominant noise contributions highlighted. The commercial seismometers' performance is given for a composite signal of a pair of \SI{1.4}{\m} separated vertical sensors. M-6D appears to have better sensitivity around the microseismic peak but this is due to the truly lower level of the microseism at the time of year when M-6D measurements were taken.}
    \label{fig:sensitivity}
\end{figure*}

The test mass motion is measured by an array of six compact interferometric sensors arranged as shown in Fig.~\ref{fig:scheme}. These consist of a customised, half-inch diameter sensing assembly derived from the SmarAct Picoscale series sensing head, a G\&H \SI{1550}{\nm} continuous-wave distributed feedback laser, and the real-time digital processing capabilities of the Control and Data acquisition System (CDS) developed for Advanced LIGO. These sensors have a demonstrated sub-pm sensitivity above \SI{0.1}{\Hz}~\cite{Smaract} and a typical non-linear deviation of $<\SI{1}{\nm}$~\cite{SmaractNonlin}. These sensors work on the principle of deep-frequency modulation (DFM)~\cite{DFM}, which involves modulating the laser frequency to a high modulation depth, thereby allocating significant optical power across multiple harmonics of the modulation frequency. By extracting the signal from at least one pair of harmonics, we can construct a linear estimator of the relative displacement between sensor and test mass. Specific details of the optical sensor, including information about the digital phase-extraction algorithm can be found in Ref.~\cite{Smaract}.

The principal benefit of interferometric sensors is through their superior sensitivity compared to typical electronic readout schemes. An additional benefit of our sensors is in the robustness of the readout algorithm during initial alignment. The sensors have an operating range spanning several centimetres. with, in principle, no impact on sensitivity. This is contingent upon choosing a laser with a sufficient modulation range to enable such large range. The sensor's linearity is invariant to angular fluctuations, with only a mild impact on sensitivity (inversely to fringe contrast, see Ref.~\cite{DFM}). Therefore, it is possible to align all the sensors in-air with relative ease. Our sensor is susceptible to strong laser frequency noise coupling. A balanced Michelson interferometer can strongly reject laser frequency noise. However, it is necessary for us to have a large arm-length imbalance to allow a strong coupling of the frequency modulation. However, as all of the sensing heads are connected to the same laser via an in-vacuum 8-way fibre splitter, the laser frequency noise is common to all sensors. Therefore, we place an extra reference sensing head inside the vacuum chamber providing a `null' measurement of a fixed mirror. This gives us an independent readout channel for the laser frequency noise, which we use in feedback to the laser frequency modulation, and thus suppress this common noise source at the laser without impacting the individual sensing heads' responses.

Mass actuation and auxiliary position sensing is achieved through the Birmingham Optical Sensors and Electromagnetic Motors (BOSEMs)~\cite{Bosem} that are symmetrically arranged similarly to the interferometric sensors. The sensing consists of a small aluminium `flag' attached to the test mass which partially obscures the light path between an LED and a photodetector creating a shadow sensor. The secondary displacement channel provided by the BOSEMs provides an estimate of the absolute test mass position over its approximate linear range of \SI{700}{\um}. This is not possible with the interferometric sensors using the current simplified algorithm due to the sinusoidal periodicity of their readout signal. The BOSEMs' nominal sensitivity is approximately 300 times worse than the interferometric sensors, so the fine position sensing during low-noise operation is entirely achieve by the latter.

The BOSEMs provide actuation to the test mass via a \SI{14.6}{\milli\henry} inductive coil coupled with a pair of cylindrical \SI{1}{\mm} by \SI{1}{\mm} samarium cobalt magnets attached to the flag. The magnets are oriented with opposing polarity to minimise the force coupling to external magnetic fields. The actuators are used as part of a sub-mHz `catching' servo, which we use to stabilise the DC mass position during initial balancing and for maintaining the same steady-state position in the face of long-term tilt drifts. We configure the actuators in one of two modes by switching the output impedance of the coil driver electronics. We use a high-range, high-noise mode during in-air stabilisation, which significantly aids with alignment of the sensors and balancing of the mass prior to pump-down. After pump-down the mass settles and we can switch to low-range, low-noise mode if necessary.

The magnets are nominally positioned at a distance from the coil such that the actuators operate at the peak gradient of the magnetic field. This is the point at which the force is maximised (as $F \propto dB/dx$) and consequently the point at which the actuator operates in the stiffness-free regime. However, this regime is difficult to sustain due to the initial difficulty in aligning all six actuators simultaneously. If a slowly-varying current with a non-zero DC component is passed through the coils while the magnets are not optimally arranged, this introduces additional stiffness, which can shift the resonant frequencies and also introduce a slow time-dependence to this shift---neither of which are desirable. This scenario occurs when the mass position is stabilised by the BOSEMs with the catching servo. Furthermore, the DC current passing through the coils induces additional eddy currents in the nearby metallic components, which can significantly reduce the Q-factors of the affected modes.

We mitigate this issue through remote balancing. The presence of the DC force applied through the catching servo is a consequence of imperfect initial balancing and subsequent tilt drifts. The neutral position differs in-air due to buoyancy effects and so a significant one-off tilt shift can be introduced after pump-down with additional gradual drifts accumulating over time. As a result, this leads to an accumulation of DC compensating force. With in-vacuum remote balancing, we are able to shift the neutral balance towards the servo set point and thus manually zero out the DC force. We achieve this by shunting a small balancing mass with a series of magnetic pulses---a system we refer to as a magnetogun. The mass consists of a \SI{3}{\mm} diameter, \SI{15}{\mm} long brass rod with a neodymium magnet bonded on each end. The balancing rods are held on the test mass by friction and the magnetic pulses are injected on the millisecond-scale. This applies a large but short-duration force to overcome the static friction and delivers an impulse to the small balancing mass without exciting the dynamics of the comparatively heavy test mass. With this technique, we are able to move the masses in approximately \SI{10}{\um} increments, corresponding to a resolution of \SI{0.5}{\micro \newton} of DC force removal for C-6D. Assuming the worst-case scenario that the steady-state tilt position will continue to drift unidirectionally by the \SI[per-mode=symbol]{2.5}{\milli \radian \per \month} observed over three months of operation indefinitely, the balanced mass will remain within range of the DC actuators for over 10 years. Given the desire to zero the DC actuation force, the mass can be periodically rebalanced with the magnetoguns. This may be at most once every few months to compensate for seasonal drifts in environmental conditions.

\section{Sensitivity}
\label{sec:sensitivity}

We present the measurements of the ground motion together with the principal noise contributions in Fig.~\ref{fig:sensitivity}. Due to the similar parameters of the C-6D mass to the metal precursor, the sensitivity of most DoFs is consistent with those already presented in Ref.~\cite{6D2}. This is the result of a similar pendulum length, moments of inertia and interferometric sensor performance. Therefore, the reader should consult the results therein, in addition to the results highlighted below, for a comprehensible overview of the 6-axis performance.

In principle, our DFM-based sensors benefit from AC readout, meaning that certain noise sources (e.g. analogue-to-digital converter noise and laser intensity noise) are down-converted from a frequency region of lower relative noise and thus bypass the characteristic increase in the low-frequency noise (see discussion in Ref.~\cite{Smaract}).

However, much like the precursor sensor, the dominant readout noise contribution comes from the nonlinearity of the sensor's readout algorithm. Both sensors ultimately obtain a linear displacement estimator by tracking the angular position along a Lissajous figure comprising two orthogonal sinusoidal signals. Despite the technical differences in obtaining the two signals, the linearity of the sensors invariably comes down to how well the parameters of this Lissajous figure can be determined. Small errors in the parameter estimation lead to periodic, nonlinear errors in the displacement estimation, which manifest themselves as an effective noise floor, which under certain circumstances can become larger than the nominal readout noise. For a more in-depth analysis and discussion of the sources of nonlinearity and their effects on sensitivity, see Ref.~\cite{SmaractNonlin}. Ultimately, we reach a similar nonlinear noise floor in these sensors compared to the metal precursor, dictated by a slight residual eccentricity to the nominally circular Lissajous figure. Here, this corresponds to an approximately 0.2\% difference between the semi-major and semi-minor axes.

The directly measured horizontal acceleration shown in Fig.~\ref{subfig:sens_x} shows real motion larger than the readout noise floor across all frequencies. We obtain a tilt-subtracted curve by performing a coherent subtraction between the `true' tilt DoF and inferred tilt from the horizontal DoF in the frequency region where the tilt-coupled motion dominates. This subtraction is straightforward owing to the simple coupling factor of $g/\omega^2$ between tilt and horizontal motion. We see a brief continuation of the true horizontal motion before reaching the sensitivity limit of the tilt DoF below \SI{50}{\mHz}. Here, the readout nonlinearities do become dominant, limiting the sensitivity at \SIrange{20}{50}{\mHz}. These are the same nonlinearities, in origin and magnitude, to those we observed with the prototype design and can be most easily seen in the Z DoF as shown in Fig. 3(b)~in Ref.~\cite{6D2}. The nonlinear noise floor is consistent with the predictions of the nonlinear noise model in fringe-counting interferometric sensors~\cite{SmaractNonlin}. The nonlinear noise can be substantially mitigated if the sensor is used in feedback within an inertial isolation scheme, as the nonlinear noise floor scales with the RMS readout signal (as shown in Ref.~\cite{SmaractNonlin}). By reducing the RMS input motion, the nonlinearities decrease as well and the nominal sub-pm sensitivity of the interferometric sensors can be realised.

By extension, the tilt sensitivity shown in Fig.~\ref{subfig:sens_tilt} is likewise dominated by the nonlinearities in this frequency band, but also contains several other significant contributions. Below \SI{10}{\mHz} and above the tilt resonances, the sensitivity is limited by a combination of actuator noise  and thermal noise, with C-6D dominated by the former and M-6D by the latter. Due to M-6D's significantly smaller moment of inertia, we could reduce the range of the tilt actuators and thus substantially reduce the actuator noise. Conversely, due to this same reduction in moment of inertia, the loss contribution from epoxy is amplified, resulting in higher thermal noise. This effect is reversed for C-6D, which leads to the actuation noise rising above the thermal noise in that case. We continue development on improved, low-noise coil-driver electronics with better frequency shaping to mitigate the actuation noise in C-6D below the thermal noise limit.

It is notable that here we present results with a lower tilt resonance than in Ref.~\cite{6D2} by approximately a factor of 2, corresponding to better sensitivity in this band. Despite the significantly smaller size of the M-6D mass, we are able to reach a similar tilt resonance for both sensors, reflecting the effectiveness of the CoM tuning in Eq.~\ref{eq:tilt_freq}. We observe the consequences of the smaller size mainly in the increased difficulty of the CoM tuning technique and the higher thermal noise.

\section{Conclusion}

We present two six-axis seismometers based on an extended fused-silica test mass suspended by a single fused-silica fibre. The purpose of these sensors is to achieve improvements in low-frequency (\unit{\mHz} to \unit{Hz}) sensitivity over existing commercial alternatives. This is in part enabled by the ability to decouple all of the degrees of freedom and thus suppress limiting cross-couplings, such as the tilt-horizontal coupling.

The sensor has great potential across a vast variety of applications from its direct use as a seismometer and also as the sensing component of a broader inertial isolation scheme. Here, we focus on its use within the Advanced LIGO active control scheme of the Internal Seismic Isolation platforms.

The two sensors presented here, C-6D and M-6D differ in size with C-6D being a factor of 2 larger. Otherwise, both sensors make use of the same sensing and actuation technology, comprising shadow sensors, compact interferometric displacement sensors, magnet-coil actuators, and pulsed magnetic mass balancers (magnetoguns). Overall, this iteration of the sensor design has a number of advantages over the metal precursors. Namely, the balancing and control of the seismometer is significantly easier, the mass is more stable and resistant to long-term drifts, and the alignment and sensing is more robust.

The key difference in performance that we see between the two sensors is in the increased difficulty of fine-tuning the resonant frequencies, the greater sensitivity to asymmetry in the 
suspension fibre profile and the consequent worse sensitivity to tilt motion for the smaller mass. However, importantly, we were able to recover or surpass the tilt sensitivity of the precursor in Ref.~\cite{6D2} despite a factor of 2.2 reduction in size for C-6D and an additional reduction by a factor of 2.5 for M-6D. In both configurations, we reach a better tilt sensitivity than a pair of Trillium 240 seismomemters, either by a factor of a few in a smaller footprint (M-6D), or by over an order of magnitude in a slightly larger footprint (C-6D).

We have tested the technology across a range of sizes, showing that it is possible to reach comparable sensitivity in either case. Therefore, we have shown that it is possible to customise the sensor's size to fit the space constraints of a given application with relatively small impact on performance and little additional R\&D.

\begin{acknowledgments}

We acknowledge members of the LIGO Suspension Working Group for useful discussions, the support of the Institute for Gravitational Wave Astronomy at the University of Birmingham, STFC Equipment Call 2018 (Grant No. ST/S002154/1), STFC Consolidated Grant ``Astrophysics at the University of Birmingham'' (No. ST/S000305/1), UKRI Quantum Technology for Fundamental Physics scheme (Grant No. ST/T006609/1 and ST/W006375/1), and UKRI ``The next-generation gravitational-wave observatory network'' project (Grant No. ST/Y00423X/1). D.M. acknowledges the support of the 2021 Philip Leverhulme Prize. 

\end{acknowledgments}

\begin{appendix}

\section*{Appendix}

\begin{figure*}
    \centering
    \begin{subfigure}{0.49\linewidth}
        \centering
        \includegraphics[width=\linewidth]{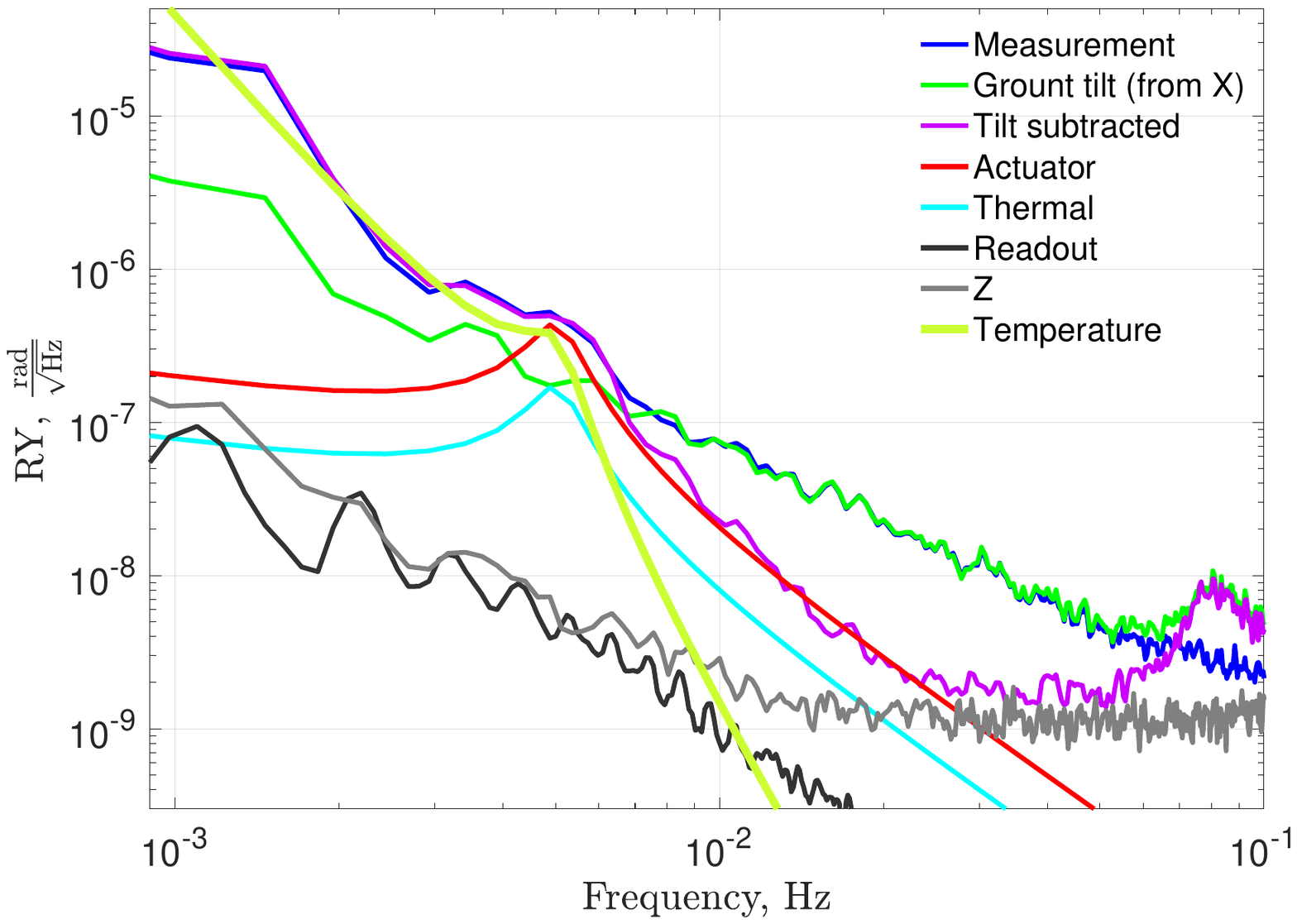}
        \caption{C-6D Tilt}
        \label{subfig:sens_c6d}
    \end{subfigure}
    \begin{subfigure}{0.49\linewidth}
        \centering
        \includegraphics[width=\linewidth]{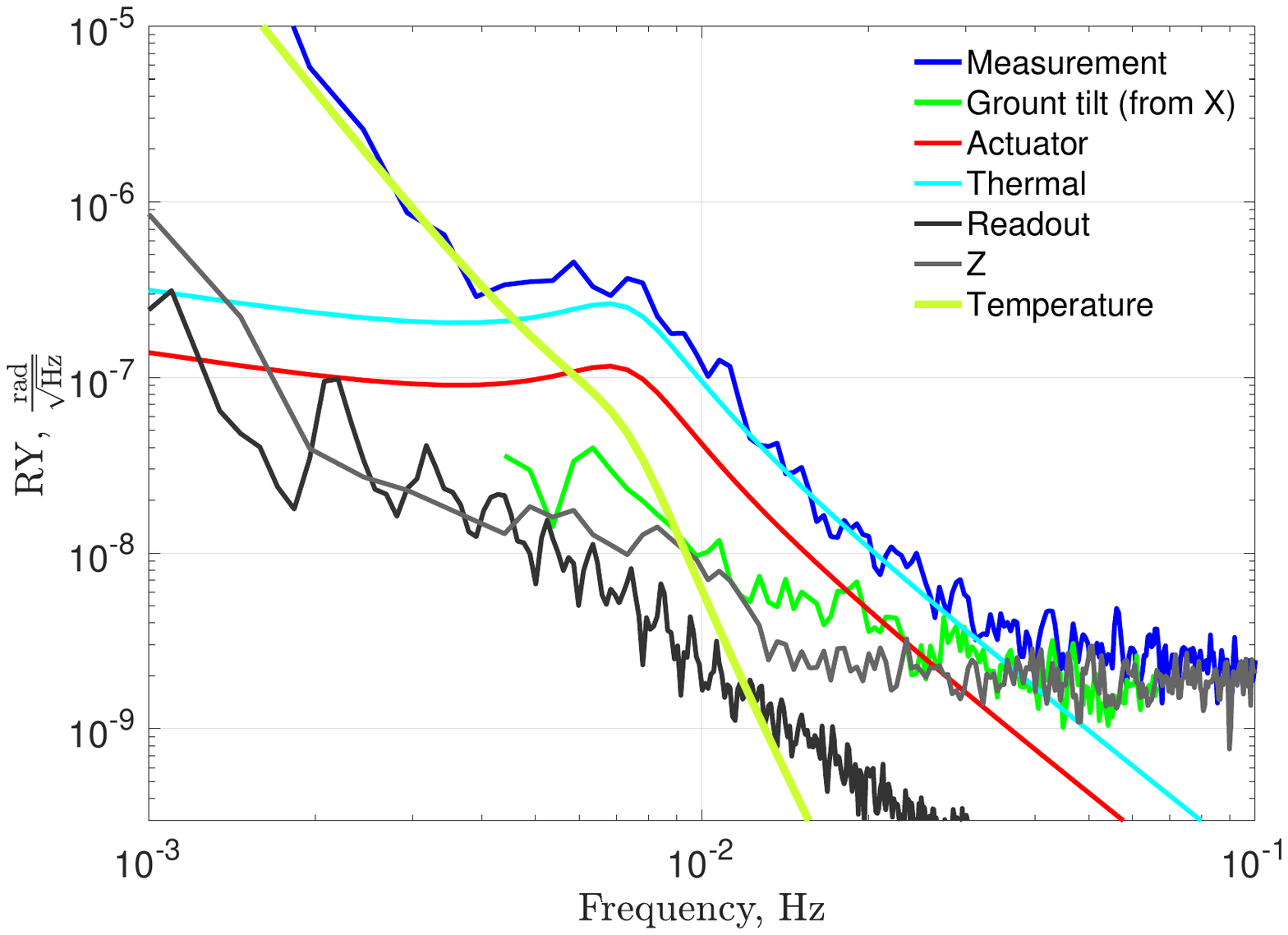}
        \caption{M-6D Tilt}
        \label{subfig:sens_m6d}
    \end{subfigure}

    \caption{Detailed tilt noise budgets of both C-6D and M-6D sensors showing the contribution from each of the key noise sources.}
    \label{fig:sens_budget}
\end{figure*}

In Fig.~\ref{fig:sensitivity} we show the final sensitivity of the two sensors. For the benefit of the interested reader, we provide the full tilt noise budget for each sensor with a breakdown of each noise contribution in Fig.~\ref{fig:sens_budget}. The measurement trace corresponds to the spectrum as measured by the vertical sensors in the nominal `tilt' DoF. The ground tilt (from X) trace corresponds to the spectrum as inferred from the cross-coupling of ground tilt into the horizontal DoF.

The tilt subtracted trace shows the result of a coherent subtraction of the horizontally-inferred tilt from the direct tilt measurement and is thus a good estimate of the residual noise below \SI{50}{\mHz}. For a better demonstration of how the tilt decoupling is performed in a full six-DoF inertial isolation of an active platform, see Ref.~\cite{6D1}. Above this frequency the subtracted signal becomes polluted by horizontal ground motion from the microseism. There is no tilt subtracted trace for M-6D as the horizontally-inferred tilt is already below the sum of all noise sources. This is due, in part, to the worse sensitivity of M-6D but also due to the naturally larger ground tilt in the winter (C-6D measurement) compared to the spring (M-6D measurement).

The individual noise sources are already discussed in Sec.~\ref{sec:sensitivity}, but it is worth pointing out the performance of the interferometric sensors. The nominal `null measurement' sensitivity of the sensors is shown in the readout trace. The nonlinear noise floor is inferred from the trace labelled `Z', which corresponds to the vertical (Z) DoF. This is the stiffest DoF and thus acts as a good estimator of the sensing noise at low frequencies. This nonlinear noise is dominant towards the upper end of the frequency band but is naturally reduced if the RMS input motion is suppressed by using the sensor in feedback. In such a case, the readout noise will cease to be the limiting noise source at any frequency.

At the lowest frequencies we are limited by temperature noise. This is due to macroscopic temperature fluctuations in the laboratory. The primary causes of this are weather conditions and the air conditioning cycle. These temperature fluctuations induce a thermal gradient across the mass, which results in very low frequency tilt fluctuations. Here, the use of fused silica is highly beneficial due to its low coefficient of thermal expansion and the only region of the sensitivity curve where the switch to fused silica can be \textit{directly} observed. This noise is estimated through an set of spatially separated in-vacuum temperature sensors which allows us to model the thermal gradients across the mass.

\end{appendix}

\bibliography{main.bib}

\end{document}